\documentclass{article}

\usepackage[scr=boondoxo,scrscaled=1.05]{mathalfa}
\usepackage[pdftex]{graphicx}
\usepackage{amsmath,amssymb,hhline,tikz,mathtools}
\usepackage{enumerate}
\usetikzlibrary{shapes.geometric,decorations.markings}
\usetikzlibrary{calc}
\usepackage[title]{appendix}
\usepackage{etoolbox}
\patchcmd{\appendices}{\quad}{: }{}{}
\usepackage{algorithm}
\usepackage{algorithmicx}
\usepackage{algpseudocode}
\usepackage{rotating}
\usepackage{afterpage}
\usepackage{xstring}
\usepackage{xcolor}
\usepackage{cmap}
\usepackage{mathdots}
\usepackage{colortbl}
\usepackage{booktabs}
\usepackage{textcomp}     
\usetikzlibrary{backgrounds}
\usepackage{pgfplots}
\pgfplotsset{compat=1.12}
\pgfmathsetseed{5112}
\usepackage{multirow}
\usepackage{pgfplotstable}
\usepackage{adjustbox}
\usepackage{diagbox}

\PassOptionsToPackage{hyphens}{url}
\usepackage{hyperref}
\hypersetup{
    pdftitle={Word Embedding Techniques for Malware Classification},
    pdfauthor={Mark Stamp},
    bookmarksnumbered=true,     
    bookmarksopen=true,         
    bookmarksopenlevel=3,       
    colorlinks=true,linkcolor=black,citecolor=black,urlcolor=blue,filecolor=blue,
    pdfstartview=Fit,           
    pdfpagemode=UseOutlines,
    pdfpagelayout=SinglePage
}

\definecolor{darkgreen}{rgb}{0.125,0.5,0.169}

\algrenewcommand{\algorithmiccomment}[1]{{\color{red}{\tt //}\ #1}}
\algnewcommand{\Initialize}[1]{%
  \State \textbf{Initialize:}
  \Statex \hspace*{\algorithmicindent}\parbox[t]{.8\linewidth}{\raggedright #1}
}
\algnewcommand{\Given}[1]{%
  \State \textbf{Given:}
  \Statex \hspace*{\algorithmicindent}\parbox[t]{.8\linewidth}{\raggedright #1}
}

\long\def\symbolfootnotetext[#1]#2{\begingroup%
\def\thefootnote{\fnsymbol{footnote}}\footnotetext[#1]{#2}\endgroup}

\let\oldsqrt\sqrt
\def\sqrt{\mathpalette\DHLhksqrt}
\def\DHLhksqrt#1#2{%
\setbox0=\hbox{$#1\oldsqrt{#2\,}$}\dimen0=\ht0
\advance\dimen0-0.2\ht0
\setbox2=\hbox{\vrule height\ht0 depth -\dimen0}%
{\box0\lower0.4pt\box2}}

\def\clap#1{\hbox to 0pt{\hss#1\hss}}

\allowdisplaybreaks

\def\figureFastSpeed{s}\def\figureSpeed{f}
\let\figureFastSpeed=\figureSpeed
\def\selectFigureSpeed#1#2{
\if\figureSpeed\figureFastSpeed #1\else #2\fi}

\def\srowvecc#1#2{(\!\begin{array}{cc} 
      \noexpandarg\IfBeginWith{#1}{-}{\! #1}{#1}
    & #2\kern-0.5pt\end{array}\!)}
\def\rowvecc#1#2{\left(\!\begin{array}{cc} 
      \noexpandarg\IfBeginWith{#1}{-}{\! #1}{#1}
    & #2\kern-0.5pt\end{array}\!\right)}
\def\rowveccc#1#2#3{\left(\!\begin{array}{ccc} 
      \noexpandarg\IfBeginWith{#1}{-}{\! #1}{#1}
    & #2 
    & #3\kern-0.5pt\end{array}\!\right)}
\def\rowvecccc#1#2#3#4{\left(\!\begin{array}{cccc}
      \noexpandarg\IfBeginWith{#1}{-}{\! #1}{#1}
    & #2 
    & #3 
    & #4\kern-0.5pt\end{array}\!\right)}
\def\srowvecccc#1#2#3#4{\bigl(\!\begin{array}{cccc}
      \noexpandarg\IfBeginWith{#1}{-}{\! #1}{#1}
    & #2 
    & #3 
    & #4\kern-0.5pt\end{array}\!\bigr)}
\def\rowveccccc#1#2#3#4#5{\left(\!\begin{array}{ccccc} 
      \noexpandarg\IfBeginWith{#1}{-}{\! #1}{#1}
    & #2
    & #3
    & #4
    & #5\kern-0.5pt\end{array}\!\right)}
\def\srowvecccccc#1#2#3#4#5#6{(\!\begin{array}{cccccc} 
      \noexpandarg\IfBeginWith{#1}{-}{\! #1}{#1}
    & #2
    & #3
    & #4
    & #5
    & #6\kern-0.5pt\end{array}\!)}
\def\rowvecccccc#1#2#3#4#5#6{\left(\!\begin{array}{cccccc} 
      \noexpandarg\IfBeginWith{#1}{-}{\! #1}{#1}
    & #2
    & #3
    & #4
    & #5
    & #6\kern-0.5pt\end{array}\!\right)}

\def\figureType{*}\def\figureSlowType{slowType}
\def\selectFigureType#1#2{
\if\figureType\figureSlowType #1\else #2\fi}
\makeatletter
\newcommand{\lowsub}[1]{\mathpalette{\raisem@th{#1}}}
\newcommand{\raisem@th}[3]{\raisebox{-#1}{$#2#3$}}
\makeatother
\def\halfthin{\kern 0.083em}

\pgfkeys{
    /pgf/number format/precision=2, 
    /pgf/number format/fixed zerofill=true }
    
\pgfplotstableset{
    /color cells/min/.initial=0,
    /color cells/max/.initial=1000,
    /color cells/textcolor/.initial=,
    color cells/.code={%
        \pgfqkeys{/color cells}{#1}%
        \pgfkeysalso{%
            postproc cell content/.code={%
                \begingroup
                \pgfkeysgetvalue{/pgfplots/table/@preprocessed cell content}\value
\ifx\value\empty
\endgroup
\else
                \pgfmathfloatparsenumber{\value}%
                \pgfmathfloattofixed{\pgfmathresult}%
                \let\value=\pgfmathresult
                \pgfplotscolormapaccess
                    [\pgfkeysvalueof{/color cells/min}:\pgfkeysvalueof{/color cells/max}]%
                    {\value}%
                    {\pgfkeysvalueof{/pgfplots/colormap name}}%
                \pgfkeysgetvalue{/pgfplots/table/@cell content}\typesetvalue
                \pgfkeysgetvalue{/color cells/textcolor}\textcolorvalue
                \toks0=\expandafter{\typesetvalue}%
                \xdef\temp{%
                    \noexpand\pgfkeysalso{%
                        @cell content={%
                            \noexpand\cellcolor[rgb]{\pgfmathresult}%
                            \noexpand\definecolor{mapped color}{rgb}{\pgfmathresult}%
                            \ifx\textcolorvalue\empty
                            \else
                                \noexpand\color{\textcolorvalue}%
                            \fi
                            \the\toks0 %
                        }%
                    }%
                }%
                \endgroup
                \temp
\fi
            }%
        }%
    }
}

\makeatletter
\newcommand*\bigcdot{\mathpalette\bigcdot@{.5}}
\newcommand*\bigcdot@[2]{\mathbin{\vcenter{\hbox{\scalebox{#2}{$\m@th#1\bullet$}}}}}
\makeatother

\def\k{\kern 2.75pt}

\newlength{\xxxxx}
\def\z{\phantom{0}}

\usepackage{epsfig}
\usepackage{multicol}
\usepackage{pslatex}
\usepackage{apalike}

\advance\oddsidemargin by -0.45in
\advance\textwidth by 0.9in
\advance\topmargin by -0.6in
\advance\textheight by 1.2in

\begin{document}

\title{Malware Classification Using Long Short-Term Memory Models}

\author{Dennis Dang\footnote{dang.dennis21@gmail.com}
\and
Fabio Di Troia\footnote{fabio.ditroia@sjsu.edu}
\and 
Mark Stamp\footnote{mark.stamp@sjsu.edu}
}

\maketitle

\abstract{Signature and anomaly based techniques are the quintessential approaches to malware detection. 
However, these techniques have become increasingly ineffective as malware has become more sophisticated 
and complex. Researchers have therefore turned to deep learning to construct better performing model. 
In this paper, we create four different long-short term memory (LSTM) based models and train each to 
classify malware samples from~20 families. Our features consist of opcodes extracted from malware 
executables. We employ techniques used in natural language processing (NLP), including word 
embedding and bidirection LSTMs (biLSTM), and we also use convolutional neural networks (CNN). 
We find that a model consisting of word embedding, biLSTMs, and CNN layers performs best in our 
malware classification experiments.}


\section{Introduction}

\subsection{Overview}
Malicious software (malware) are computer programs that are created to harm a computer, 
computer systems or a computer user~\cite{malware_background1}. Malware attacks can disrupt a person's 
or organization's day-to-day use of their computer systems, steal personal or confidential information, 
corrupt files or annoy users. Malware can be categorized into different families where the behavior of 
malware from one particular family differs from that of another family. The papers~\cite{malware_background2} 
and~\cite{pratik_paper}, for example, discuss the behavior of many different malware families. 

Modern malware attacks are generally facilitated by the Internet. With the rise in the number of devices 
that are connected to the Internet, it has become more important than ever to keep our devices safe, 
lest we risk loss of personal or confidential information~\cite{malware_background2}. While many
malware attacks are often annoying, some 
can be life threatening. An example of the latter occurred In 2017
when a ransomware\footnote{Ransomware is a type of malware that 
threatens to corrupt, delete, publish or block the victim's data unless a ransom is paid.} 
attack crippled parts the United Kingdom's National Health Service (NHS)~\cite{ransomware_attack}. 
Computer systems containing data pertaining to the health of thousands of patients were 
targeted across dozens of hospitals in the UK. Hospitals were forced to pay a ransom to have their files 
unlocked or risk having their files corrupted or deleted. These attacks caused doctors 
and nurses to cancel some~19,000 appointments, and they cost the NHS~£92 million. 
Malware is clearly a security challenge that warrants a significant research effort.

Malware detection techniques include signature based detection, anomaly based detection, 
and machine learning based detection~\cite{malware_background1}. Signature based detection has long
been the most popular approach to detecting malware. In a signature based approach, each malware sample 
is first analyzed and a signature is extracted, which is then used to identify the malware. A signature is
typically a carefully chosen, fixed bit string that is extracted from a malware sample. If the 
signature is found in another sample, that sample is flagged as possible malware. 
However, various code obfuscation and code morphing techniques can easily thwart signature based 
detection mechanisms. 

An anomaly based detection system looks for activity that falls outside 
the ``normal'' range of a computer~\cite{anomaly_detection}, and such behavior is flagged as 
suspicious. Anomaly based systems often suffer from a high false positive rate. The
drawbacks of signature and anomaly based detection has motivated the 
rise of machine learning techmiques.

Many classical machine learning algorithms have found success in detecting malware~\cite{classical_ml}. 
These algorithms include support vector machines (SVM), hidden markov models (HMM), random forest, 
and naive Bayes, among many others. Such models rely heavily on proper feature extraction from the 
dataset. Deep learning techniques have also gained considerable traction---multilayer perceptrons (MLP), 
convolutional neural networks (CNN), and extreme learning machines (ELM) have all been used 
with success~\cite{mugda_paper}. Other techniques involving variants of recurrent neural networks (RNN),
such as gated recurrent units (GRU) and long-short term memory (LSTM) models have received far less 
attention in the literature~\cite{lstm_opcodes}. 

In this research, we focus on using LSTMs to classify malware by family.
We build on the work in~\cite{lstm_opcodes} by combining various aspects of the
methodologies employed in~\cite{lstm_gru_cnn},~\cite{lstm_cnn}, and~\cite{lstm_cnn_cloud}. 
Our dataset includes malware belonging to~20 distinct families, and we use opcode sequences as
our features. We consider five models, with each model being successively more complex. 
Our first model is the most basic consisting of only MLPs. This model serves as a baseline from which we 
compare our other LSTM models to. 
Our second model consists of only one LSTM layer. Our third model is an enhanced
LSTM that includes an embedding layer, similar to the model considered in~\cite{lstm_opcodes}. 
Our fourth model replaces the LSTM layer from our second previous model with a biLSTM layer. 
Finally, our fifth model includes everything from our third model, plus an additional 
one-dimension CNN layer and a one-dimension max pooling layer. As far as we are a aware,
our fourth and fifth models have not previously been considered in the literature.

The remainder of this paper is organized as follows. Section~\ref{sect:2} discusses 
relevant previous work and introduces the various deep learning techniques employed in this 
research. Section~\ref{sect:3} covers the dataset, feature extraction, parameters,
and so on. In Section~\ref{sect:4}, we present our experimental results. Finally, 
Section~\ref{sect:5} concludes the paper, and we mention possible directions for 
future work.

\section{Background}\label{sect:2}

\subsection{Related Work}
The authors of~\cite{lstm_gru_cnn} consider various models for malware classification. 
In one of these models, a two stage classifier is used---the first stage is either 
an LSTM or GRU which is used to derive features for a second stage classifier consisting of a single MLP layer. 
Another model uses a single stage classifier consisting of nine CNN layers. When trained and evaluated, 
both models achieved an about~80\%\ accuracy.

In~\cite{lstm_cnn}, the author proposes a novel deep learning architecture 
that includes both a CNN layer and an LSTM layer. This model is trained on API call 
sequences. The CNN portion of the model consists of filters of increasing size,
with the output of each filter fed into the LSTM layer. The output of the LSTM layer is 
used as input to a dropout layer, with a final fully connected layer for classification. 
The output of the dense layer is the model's prediction for the given input. 
This model achieved an accuracy approaching~100\%.  

The authors of~\cite{lstm_cnn_cloud} consider a biLSTM based model to classify malware in 
a cloud-based system. The model includes a CNN layer and is trained on system call sequences. 
The authors achieve an overall accuracy of approximately~90\%. Interestingly, the authors also 
show that substituting the biLSTM for a regular LSTM layer resulted in worse accuracies in almost all cases.

The author in~\cite{lstm_opcodes} classifies malware using an entirely different approach from 
the two papers mentioned above. The work in~\cite{lstm_opcodes} is based on opcodes obtained 
from disassembled executables. This research also employs word embedding 
as a feature engineering step. Word embedding techniques are often used in 
natural language processing (NLP) applications. The result from word embedding are 
fed into an LSTM layer. For malware detection, this model attains an average AUC 
of~0.99, while for classification, the model achieves an average AUC of~0.987.

\subsection{Recurrent Neural Networks}
In feedforward neural networks, all training samples are treated independently of each 
other~\cite{professor_stamps_great_novel}. Consequently, feedforward networks are
impractical for cases where training samples depend on previous samples. Thus, a 
different type of architecture is needed in cases where ``memory'' is required,
as when training on time series or other sequential data.

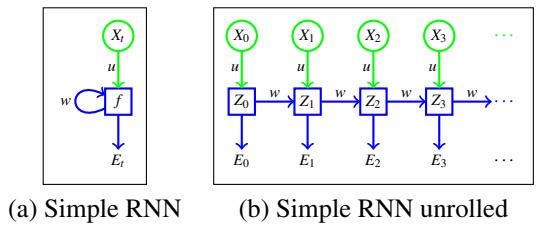
\begin{figure}[!htb]
\centering
  \begin{tabular}{cc}
  \begin{tikzpicture}[framed,scale=0.35,every node/.style={scale=0.6}]


    

    


    \draw[thick,color=green] (10.5,5.5) circle (0.575);
    \draw[thick,color=blue] (10.0,2.5) rectangle (11.0,3.5);
    \draw[thick,color=green,->] (10.5,4.925) -- (10.5,3.5);
    \draw[thick,color=blue,->] (10.5,2.5) -- (10.5,1.17);

    \draw[thick,color=blue,->] (10.0,2.75) .. controls (8.5,2.0) and (8.5,4.0) .. (10.0,3.25);

    \node at (8.5,3.0) {$w$};
    \node at (10.25,4.3) {$u$};

    \node at (10.5,5.5) {$X_{t}$};
    \node at (10.5,3.0) {$f$};
    \node at (10.5,0.75) {$E_{t}$};

\end{tikzpicture}
  &
  \begin{tikzpicture}[framed,scale=0.35,every node/.style={scale=0.6}]

    \draw[thick,color=green] (-2.0,5.5) circle (0.575);
    \draw[thick,color=blue] (-2.5,2.5) rectangle (-1.5,3.5);
    \draw[thick,color=green,->] (-2.0,4.925) -- (-2.0,3.5);
    \draw[thick,color=blue,->] (-2.0,2.5) -- (-2.0,1.17);
    \draw[thick,color=blue,->] (-1.5,3.0) -- (0.0,3.0);

    \node at (-2.25,4.3) {$u$};
    \node at (-2.0,5.5) {$X_{0}$};
    \node at (-2.0,3.0) {$Z_{0}$};
    \node at (-2.0,0.75) {$E_{0}$};

    \draw[thick,color=green] (0.5,5.5) circle (0.575);
    \draw[thick,color=blue] (0.0,2.5) rectangle (1.0,3.5);
    \draw[thick,color=green,->] (0.5,4.925) -- (0.5,3.5);
    \draw[thick,color=blue,->] (0.5,2.5) -- (0.5,1.17);
    \draw[thick,color=blue,->] (1.0,3.0) -- (2.5,3.0);

    \node at (0.25,4.3) {$u$};
    \node at (-0.75,3.3) {$w$};
    \node at (0.5,5.5) {$X_{1}$};
    \node at (0.5,3.0) {$Z_{1}$};
    \node at (0.5,0.75) {$E_{1}$};
    
    \draw[thick,color=green] (3.0,5.5) circle (0.575);
    \draw[thick,color=blue] (2.5,2.5) rectangle (3.5,3.5);
    \draw[thick,color=green,->] (3.0,4.925) -- (3.0,3.5);
    \draw[thick,color=blue,->] (3.0,2.5) -- (3.0,1.17);
    \draw[thick,color=blue,->] (3.5,3.0) -- (5.0,3.0);

    \node at (2.75,4.3) {$u$};
    \node at (1.75,3.3) {$w$};
    \node at (3.0,5.5) {$X_{2}$};
    \node at (3.0,3.0) {$Z_{2}$};
    \node at (3.0,0.75) {$E_{2}$};
    
    \draw[thick,color=green] (5.5,5.5) circle (0.575);
    \draw[thick,color=blue] (5.0,2.5) rectangle (6.0,3.5);
    \draw[thick,color=green,->] (5.5,4.925) -- (5.5,3.5);
    \draw[thick,color=blue,->] (5.5,2.5) -- (5.5,1.17);
    \draw[thick,color=blue,->] (6.0,3.0) -- (7.5,3.0);

    \node at (5.25,4.3) {$u$};
    \node at (4.25,3.3) {$w$};
    \node at (5.5,5.5) {$X_{3}$};
    \node at (5.5,3.0) {$Z_{3}$};
    \node at (5.5,0.75) {$E_{3}$};

    \node at (6.75,3.3) {$w$};

    \node[color=green] at (8.0,5.5) {\large$\cdots$};
    \node[color=blue] at (8.0,3.0) {\large$\cdots$};
    \node[color=black] at (8.0,0.75) {\large$\cdots$};
    
\end{tikzpicture}
  \\
  (a) Simple RNN
  &
  (b) Simple RNN unrolled
  \\[2ex]
  \end{tabular}
  \caption{Simple RNN and its unrolled version}\label{fig:RNN_simp_unroll}
\end{figure}

Recurrent neural networks (RNN) serve to add memory to the network~\cite{rnn_background}. 
As illustrated in Figure~\ref{fig:RNN_simp_unroll}~(a), the output in a RNN depends not only on the current
input, but also the input from the past, as indicated by a feedback loop. Whereas information 
only flows forward in a feed-forward network, information from the previous timesteps are 
available at each subsequent timestep in RNNs~\cite{rnn_ann}. An unrolled view of an RNN~\cite{rnn_unrolled} 
appears in Figure~\ref{fig:RNN_simp_unroll}~(b).

\subsection{Long Short-Term Memory}
While conceptually simple, plain vanilla RNNs suffer from the ``vanishing gradient'' issue when training
via backpropagation, which severely limits the ``memory'' available to the model. 
To overcome this gradient issue, complex gated RNN architectures have been 
developed---the best known and most widely used of these is long short-term memory (LSTM) 
models.  
LSTMs address the issue of long-term dependency by, in effect,
decoupling the memory from the output of the network and ensuring that additive 
updates are done to the memory, rather than multiplicative updates. With additive updates, 
the gradient is more stable.

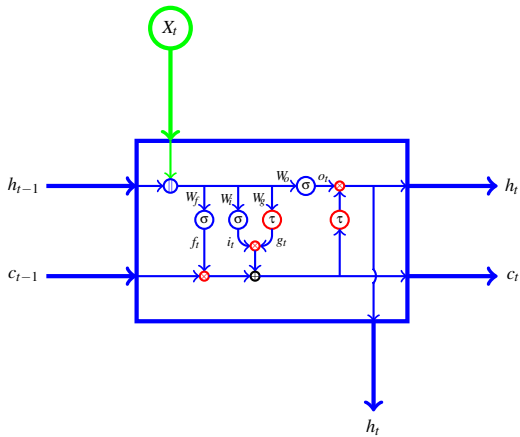
\begin{figure}[!htb]
\centering
  \begin{tikzpicture}[scale=0.6,every node/.style={scale=0.7}]

    \draw[ultra thick,color=blue] (0.0,0.0) rectangle (6.0,4.0);
    \draw[ultra thick,color=green] (0.75,6.5) circle(0.45);
    \draw[ultra thick,color=green,->] (0.75,6.05) -- (0.75,4.0);
    \draw[ultra thick,color=blue,->] (5.25,0.0) -- (5.25,-2.0);
    \draw[ultra thick,color=blue,->] (-2.0,1.0) -- (0.0,1.0);
    \draw[ultra thick,color=blue,->] (-2.0,3.0) -- (0.0,3.0);
    \draw[ultra thick,color=blue,->] (6.0,1.0) -- (8.0,1.0);
    \draw[ultra thick,color=blue,->] (6.0,3.0) -- (8.0,3.0);

    \node at (-2.5,1.0) {$c_{t-1}$};
    \node at (-2.5,3.0) {$h_{t-1}$};
    \node at (0.75,6.5) {$X_{t}$};
    \node at (5.25,-2.35) {$h_{t}$};
    \node at (8.35,1.0) {$c_{t}$};
    \node at (8.325,3.0) {$h_{t}$};


    \draw[thick,color=green,->] (0.75,4.0) -- (0.75,3.15);
    \draw[thick,color=blue] (0.75,3.0) circle(0.15);
    \node at (0.75,3.0) {$\color{blue}\scriptstyle\|$};

    \draw[thick,color=blue,->] (0.0,3.0) -- (0.60,3.0);
    \draw[thick,color=blue,->] (0.90,3.0) -- (3.55,3.0);
    \draw[thick,color=blue] (3.75,3.0) circle(0.2);
    \node at (3.75,3.0) {$\scriptstyle\sigma$};
    \node at (3.25,3.2) {$\scriptstyle W_{\kern -1.5pt o}$};
    \draw[thick,color=blue,->] (3.95,3.0) -- (4.4,3.0);
    \node at (4.15,3.175) {$\scriptstyle o_t$};


    \draw[thick,color=blue,->] (1.5,3.0) -- (1.5,2.45);
    \draw[thick,color=blue] (1.5,2.25) circle(0.2);
    \node at (1.5,2.25) {$\scriptstyle\sigma$};
    \node at (1.25,2.7) {$\scriptstyle W_{\kern -1.5pt f}$};
    \draw[thick,color=blue,->] (1.5,2.05) -- (1.5,1.1);
    \node at (1.3,1.75) {$\scriptstyle f_t$};
    \draw[thick,color=red] (1.5,1.0) circle(0.1);
    \node at (1.5,1.0) {$\color{red}\scriptstyle\times$};         
 
    \draw[thick,color=blue,->] (2.25,3.0) -- (2.25,2.45);
    \draw[thick,color=blue] (2.25,2.25) circle(0.2);
    \node at (2.0,2.7) {$\scriptstyle W_{\kern -1.0pt i}$};
    \node at (2.25,2.25) {$\scriptstyle\sigma$};    
    \draw[thick,color=blue,rounded corners,->] (2.25,2.05) -- (2.25,1.67) -- (2.525,1.67);
    \node at (2.1,1.75) {$\scriptstyle i_t$};

    \draw[thick,color=blue,->] (3.0,3.0) -- (3.0,2.45);
    \node at (2.725,2.7) {$\scriptstyle W_{\kern -1.5pt g}$};
    \draw[thick,color=red] (3.0,2.25) circle(0.2);
    \node at (3.0,2.25) {$\scriptstyle\tau$};     
    \draw[thick,color=blue,rounded corners,->] (3.0,2.05) -- (3.0,1.67) -- (2.725,1.67);
    \draw[thick,color=red] (2.625,1.67) circle(0.1);
    \node at (2.625,1.67) {$\color{red}\scriptstyle\times$};         
    \draw[thick,color=blue,->] (2.625,1.57) -- (2.625,1.1);
    \draw[thick,color=black] (2.625,1.0) circle(0.1);
    \node at (2.625,1.0) {$\scriptstyle\hbox{}+\hbox{}$};         
    \node at (3.225,1.75) {$\scriptstyle g_t$};


    \draw[thick,color=blue,->] (4.5,2.45) -- (4.5,2.9);
    \draw[thick,color=red] (4.5,2.25) circle(0.2);
    \node at (4.5,2.25) {$\scriptstyle\tau$};    
    \draw[thick,color=blue,->] (4.5,1.0) -- (4.5,2.05);

    \draw[thick,color=blue,->] (4.6,3.0) -- (6.0,3.0);
    \draw[thick,color=red] (4.5,3.0) circle(0.1);
    \node at (4.5,3.0) {$\color{red}\scriptstyle\times$};         

    \draw[thick,color=blue,->] (0.0,1.0) -- (1.4,1.0);
    \draw[thick,color=blue,->] (1.6,1.0) -- (2.525,1.0);
    \draw[thick,color=blue,->] (2.725,1.0) -- (6.0,1.0);

    \draw[thick,color=blue] (5.25,3.0) -- (5.25,1.15);
    \draw[thick,color=blue,->] (5.25,0.85) -- (5.25,0.0);
    \draw[thick,color=blue,rounded corners] (5.25,1.15) -- (5.4,1.0) -- (5.25,0.85);
    
\end{tikzpicture}
\caption{One timestep of an LSTM}\label{fig:LSTM_1}
\end{figure}

One timestep of an LSTM is illustrated in Figure~\ref{fig:LSTM_1}. The cell state~$c_t$
serves as a repository for long term memory that can be tapped when needed. 
The ``gate'' represented by~$W_{\kern -1.5pt f}$ enables the model to ``forget'' 
information in the cell state, ~$W_{\kern -1.0pt i}$ and~$W_{\kern -1.5pt g}$ 
together serve to add ``memory'' to the cell state, and the 
structure involving the output gate~$W_{\kern -1.5pt o}$
allows the model to draw on the stored memory in the cell state.

A detailed discussion of LSTMs is beyond the scope of this paper. For more information
on LSTMs, see~\cite{lstm_intro}, for example.

\subsection{Bidirectional LSTM}
BiLSTM models are an extension of LSTMs that process a sequence of data in both forward and backward 
directions in two separate LSTM layers. The forward layer processes 
the data in the same way as a standard LSTM, while the backward layer processes the same data but in reverse order~\cite{bidirectional_lstm}. As with LSTMs, a detailed discussion of biLSTMs is beyond the scope
of this paper---see, for example,~\cite{bidirectional_lstm_more_indepth} for more details.

\subsection{Word2Vec}
Word2Vec is a technique for embedding ``words'' into a
high-dimensional space. These word embeddings
are obtained by training a shallow neural network.
After the training process, 
words that are more similar in context will tend to be 
closer together in the Word2Vec space.

Perhaps surprisingly, meaningful algebraic properties also hold for Word2Vec
embeddings. For example, according to~\cite{w2v}, if we let
$$
  w_0=\mbox{\footnotesize ``king''}, w_1=\mbox{\footnotesize ``man''}, 
  w_2=\mbox{\footnotesize ``woman''}, w_3=\mbox{\footnotesize ``queen''}
$$
and~$V(w_i)$ is the Word2Vec embedding of word~$w_i$, then~$V(w_3)$
is the vector that is closest---in terms of cosine similarity---to
$$
  V(w_0) - V(w_1) + V(w_2)
$$
Results such as this indicate that Word2Vec embeddings of English text
capture significant aspects of the semantics of the language.

In the context of this paper, the ``words'' are mnemonic opcodes.
We use Word2Vec embeddings as form of feature engineering, with
the Word2Vec vectors serving as input features to our models.
Previous research has shown that Word2Vec features
are more informative than raw opcode features~\cite{aniket_paper}.

\subsection{Convolutional Neural Networks}
Convolutional neural networks (CNNs) are designed primarily to efficiently
deal with local structure~\cite{stamp_supplment}. CNNs were originally designed
for use in image classification, but the technique is applicable in any situation where 
some form of local structure dominates.

The hidden layers within a CNN act as filters where each filter specializes in 
detecting a certain feature within the data, while deeper layers detect progressively
more abstract features. For example, when training on images, 
first layer filters might detect vertical and horizontal lines, 
the final layer might be able to distinguish between images of, say, dogs
and cats.

While not strictly required, pooling layers can be applied in between CNN layers. 
These layers reduce the dimensionality, thereby reducing the computational load. Pooling
can also reduce noise and potentially improve performance. In max pooling, we specify a window size 
and only the maximum value within each (non-overlapping) window is retained.

\subsection{TensorFlow Layers}
TensorFlow models are created by adding various layers in sequence. What distinguishes one model from 
another is the type of layers used and the parameters passed into the constructors of each layer. 
A short description of each layer is provided below~\cite{tensorflow_API}.

 \begin{itemize}
     \item \textbf{Input Layer}: 
     The first layer and entry point into a neural network
     
     \item \textbf{Dropout Layer}: 
     Adds noise to the network during training by randomly severing the number of connections between 
     neurons from one layer to the next. In doing so, overfitting is reduced allowing models to better generalize. 
     This typically has the effect of increasing model accuracy during evaluation.
     
     \item \textbf{LSTM Layer}: 
     Implements a single LSTM layer with all of the algorithms required for forward and backward propagation.
     
     \item \textbf{Bidirectional Layer}: 
     A wrapper layer that allows RNN layers to implement bidirectional models. Rather than implementing two 
     separate RNN layers for the forwards and backwards direction and concatenating the results, 
     the bidirectional wrapper layer does all of this in one layer.
     
     \item \textbf{Dense Layer}:
     Implements a single fully connected vanilla neural network layer.
     
     \item \textbf{Embedding Layer}: 
     Responsible for mapping positive integers to vectors of floating point values. 
     
     \item \textbf{Conv1D Layer}:
     Implements the convolutional neural network layer in one dimension.
     
     \item \textbf{MaxPooling1D Layer}:
     Implements the max pooling operation in one dimension.
 \end{itemize}

\section{Dataset and Experimental Design}\label{sect:3}

The dataset used in this research was acquired from~\cite{pratik_paper} and from~\cite{malicia_dataset}. 
Our dataset consists of binary files from~20 distinct malware families. The names of the malware families 
and the number of samples per family is shown in Table~\ref{tab:num_malware_per_family}. 

To extract features from our dataset, we first disassemble every executable file and extracted mnemonic opcode sequences. Afterwards, we perform a frequency analysis on all opcodes. The results from this frequency analysis is used to sort opcodes in order of decreasing frequency. Next, we create an opcode to integer mapping where each opcode is assigned a unique integer, Finally, we use this mapping to convert each opcode mnemonic into integers.

We retain the~30 most frequent opcodes, with all remaining opcodes grouped into a single ``other'' category.
Each omitted opcode contributes less than~0.5\%\ to the total number of opcodes an hence would have
minimal effect on sequence-based techniques. Note that this approach has been used many recent studies, including~\cite{aniket_paper,mugda_paper,pratik_paper}.

\begin{table}[!htb]
\centering
\caption{Number of samples per malware family}
\label{tab:num_malware_per_family}
\resizebox{0.225\textwidth}{!}{
\begin{tabular}{@{}c|c@{}}
\toprule
\multicolumn{1}{c|}{\textbf{Malware Family}} & \textbf{Samples} \\
\midrule \midrule
Adload & 1044 \\
Agent & \z817 \\
Alureon & 1327 \\
BHO & 1159 \\
CeeInject & \z886 \\
Cycbot & 1029 \\
DelfInject & 1097 \\
Fakerean & 1063 \\
Hotbar & 1476 \\
Lolyda & \z915 \\
Obfuscator & 1331 \\
Onlinegames & 1284 \\
Rbot & \z817 \\
Renos & 1309 \\
Starpage & 1084 \\
Vobfus & \z924 \\
Vundo & 1784 \\
Winwebsec & 3651 \\
Zbot & 1785 \\
Zeroacess & 1119 \\ \midrule
Total & 25,901\z \\ \bottomrule
\end{tabular}
}
\end{table}

The models used in this research require all input data to be of the same length. 
To accomplish this, we experimented with various opcode sequence lengths, as discussed below.
Of course, truncating the opcode sequence results in a loss of information, but using a short sequence improves
efficiency. Our results show that we can obtain strong results with relatively short opcode sequences.

\subsection{Hardware and Software}
The models used in this research were run on a PC desktop. The specifications of 
this machine is shown in Table~\ref{tab:hard_specs}. In addition, the software, 
operating system, and Python packages used are specified in Table~\ref{tab:software_specs}.

\begin{table}[!htb]
\centering
\caption{Relevant hardware specifications}
\label{tab:hard_specs}
\resizebox{0.475\textwidth}{!}{
\begin{tabular}{@{}cll@{}}
\toprule
\multicolumn{1}{c}{\textbf{Hardware}} & \multicolumn{1}{c}{\textbf{Feature}} & \multicolumn{1}{c}{\textbf{Details}} \\
\midrule \midrule
\multicolumn{1}{c|}{\multirow{4}{*}{CPU}} & \multicolumn{1}{l|}{Brand and Model} & Intel i7-8700 \\
\multicolumn{1}{c|}{} & \multicolumn{1}{l|}{Base Clock Speed} & 3.2 GHz \\
\multicolumn{1}{c|}{} & \multicolumn{1}{l|}{\# Core} & 6 \\
\multicolumn{1}{c|}{} & \multicolumn{1}{l|}{\# Threads} & 12 \\ \midrule
\multicolumn{1}{c|}{\multirow{4}{*}{GPU}} & \multicolumn{1}{l|}{Chipset} & NVIDIA GeForce GTX 1070 Ti \\
\multicolumn{1}{c|}{} & \multicolumn{1}{l|}{Video Memory} & 8GB GDDR5 \\
\multicolumn{1}{c|}{} & \multicolumn{1}{l|}{Memory Speed} & 1683 MHz \\
\multicolumn{1}{c|}{} & \multicolumn{1}{l|}{Cuda Cores} & 2432 \\ \midrule
\multicolumn{1}{c|}{\multirow{3}{*}{DRAM}} & \multicolumn{1}{l|}{Brand and Model} & G. Skill TridentZ RGB Series \\
\multicolumn{1}{c|}{} & \multicolumn{1}{l|}{Amount} & 2 $\times 8\text{GB} = 16$GB \\
\multicolumn{1}{c|}{} & \multicolumn{1}{l|}{Speed} & 3200MHz \\ \midrule
\multicolumn{1}{c|}{Motherboard} & \multicolumn{1}{l|}{Brand and Model} & MSI Z370 SLI Plus LGA 1151 \\ \bottomrule
\end{tabular}
}
\end{table}

\begin{table}[!htb]
\centering
\caption{Relevant software, operating system, and Python packages}
\label{tab:software_specs}
\resizebox{0.4\textwidth}{!}{
\begin{tabular}{@{}cc@{}}
\toprule
\textbf{Software} & \textbf{Version} \\ 
\midrule \midrule
\multicolumn{1}{c|}{OS} & Windows 10 Pro \\
\multicolumn{1}{c|}{Python} & 3.8.3 \\
\multicolumn{1}{c|}{Jupyter Notebook} & 6.1.4 \\
\multicolumn{1}{c|}{Numpy} & 1.18.5 \\
\multicolumn{1}{c|}{Scikit Learn} & 0.23.2 \\
\multicolumn{1}{c|}{Tensorflow-GPU} & 2.3.1 \\
\multicolumn{1}{c|}{CUDA Toolkit} & 10.1 \\
\multicolumn{1}{c|}{cuDNN SDK} & 7.6 \\
\multicolumn{1}{c|}{NVidia GPU Drivers} & 431.36 \\
\multicolumn{1}{c|}{Oracle VM VirtualBox} & 6.0.10 \\
\multicolumn{1}{c|}{VM OS} & Ubuntu 18.04.5 LTS \\ \bottomrule
\end{tabular}
}
\end{table}

\subsection{Model Parameters}
Deep learning models generally have many parameters that require tuning. 
For each of our models, we performed
a grid search over reasonable values for a wide range of parameters---all combinations 
of the values tested are listed in Table~\ref{tab: grid_search_values}.
All models were trained and evaluated on the same dataset. 
For every model evaluated, 
the accuracy was determined and the parameters for the model with highest
accuracy were generally selected. In a few cases where accuracy differences were 
deemed insignificant, we selected parameters so that training times were reduced.
In Table~\ref{tab:model_parameters}, we list the specific values of the parameters 
that were selected. These parameter were used for all subsequent experiments
considered in this paper.

\begin{table}[!htb]
\centering
\caption{Parameters tested}
\label{tab: grid_search_values}
\resizebox{0.475\textwidth}{!}{
\begin{tabular}{@{}ll@{}}
\toprule
\textbf{Parameter} & \textbf{Values Tested} \\ 
\hline \hline
Opcode Lengths & {[}2000, 4000, 6000, 8000, 10000{]} \\
\midrule
LSTM Units & {[}16, 32, 64, 128, 256{]} \\
\midrule
Embedding Vector Lengths & {[}16, 32, 64, 128, 256{]} \\
\midrule
Dropout Amount & {[}0.1, 0.2, 0.3, 0.4{]} \\ \bottomrule
\end{tabular}
}
\end{table}

\begin{table}[!htb]
\centering
\caption{Parameters selected}
\label{tab:model_parameters}
\resizebox{0.375\textwidth}{!}{
\begin{tabular}{@{}lc@{}}
\toprule
\textbf{Parameter} & \textbf{Value} \\ 
\hline \hline
Batch Size & 32 \\
\midrule
Maximum Number of Epochs & 100 \\
\midrule
Percentage of Data to be Used in Testing & 15\% \\
\midrule
Number of  Unique Opcodes Used & 30 \\
\midrule
Opcode Sequence Length & 2000 \\
\midrule
Dropout Amount & 30\% \\
\midrule
Number of LSTM Units & 16 \\
\midrule
Embedding Vector Length & 128 \\
\midrule
CNN Kernal Size & 3 \\
\midrule
Number of CNN Filters & 128 \\
\midrule
Max Pooling Size & 2 \\ \bottomrule
\end{tabular}
}
\end{table}

\subsection{Training and Testing}
The dataset was sorted in ascending order based on the number of training samples per family.
The dataset was then partitioned into four groups of five families each, 
where the first group consisted of families with the most malware samples, 
while the last group consisted of families with the least samples. The models were trained on 
the first group of~5 families, then the second group of~10 (i.e., the first and second groups of~5), 
then the third group of~15, and finally on all families together. With each additional group, 
the difficulty of classifying malware by family increased---not only due to the inherent 
difficulty of having more classes, but also due to more limited training data
for some of the families. Table~\ref{tab:data_groups} lists
the families that constitute each group, while Table~\ref{tab:files_per_group} gives the number of 
training and testing samples for each group considered.

\begin{table}[!htb]
\centering
\caption{Groupings of families}
\label{tab:data_groups}
\resizebox{0.225\textwidth}{!}{
\begin{tabular}{@{}cc@{}}
\toprule
\textbf{Group} & \textbf{Malware Families}\\
\midrule \midrule
\multicolumn{1}{c|}{\multirow{5}{*}{1}} & Hotbar \\
\multicolumn{1}{c|}{} & Renos \\
\multicolumn{1}{c|}{} & Vundo \\
\multicolumn{1}{c|}{} & Winwebsec \\
\multicolumn{1}{c|}{} & Zbot \\ \midrule
\multicolumn{1}{c|}{\multirow{5}{*}{2}} & Alureon \\
\multicolumn{1}{c|}{} & Bho \\
\multicolumn{1}{c|}{} & Obfuscator \\
\multicolumn{1}{c|}{} & Onlinegames \\
\multicolumn{1}{c|}{} & Zeroaccess \\ \midrule
\multicolumn{1}{c|}{\multirow{5}{*}{3}} & Adload \\
\multicolumn{1}{c|}{} & Cycbot \\
\multicolumn{1}{c|}{} & Delfinject \\
\multicolumn{1}{c|}{} & Fakerean \\
\multicolumn{1}{c|}{} & Startpage \\ \midrule
\multicolumn{1}{c|}{\multirow{5}{*}{4}} & Agent \\
\multicolumn{1}{c|}{} & Ceeinject \\
\multicolumn{1}{c|}{} & Lolyda \\
\multicolumn{1}{c|}{} & Rbot \\
\multicolumn{1}{c|}{} & Vobfus \\ \bottomrule
\end{tabular}
}
\end{table}

\begin{table}[!htb]
\centering
\caption{Number of samples for training and testing}
\label{tab:files_per_group}
\resizebox{0.35\textwidth}{!}{
\begin{tabular}{@{}ccrr@{}}
\toprule
\multirow{2}{*}{\textbf{Groups}} & \multirow{2}{*}{\textbf{Families}} &
\multicolumn{2}{c}{\textbf{Samples}} \\
	& & \textbf{Training} & \textbf{Testing} \\ 
\midrule \midrule
1          &   \z5 & 8480 & 1472 \\
1,2       & 10 & 13,760 & 2400 \\
1,2,3    & 15 &  18,272 & 3200 \\
1,2,3,4 & 20 &  21,984 & 3872 \\ \bottomrule
\end{tabular}
}
\end{table}

The initial values of the weights of the LSTM are randomly selected
and the embedding and dense layers are randomly initialized each time the models are trained.
As a result of this random initialization, the model will likely differ, and hence the accuracy will also
likely vary each time a model is trained. 
Therefore, we train each model type
on each grouping of malware families five times. At the start of every training run, 
the dataset is shuffled before being split into training and testing sets. 
The average of these five cases is used to compare the different model types. 

\section{Experiments and Results}\label{sect:4}

In this section, we give experimental results for each of the four model
types tested. We conclude this section with a comparison of the different models. 

\subsection{Using MLP Only}

\begin{figure}[!htb]
    \centering
    \includegraphics[scale=0.3]{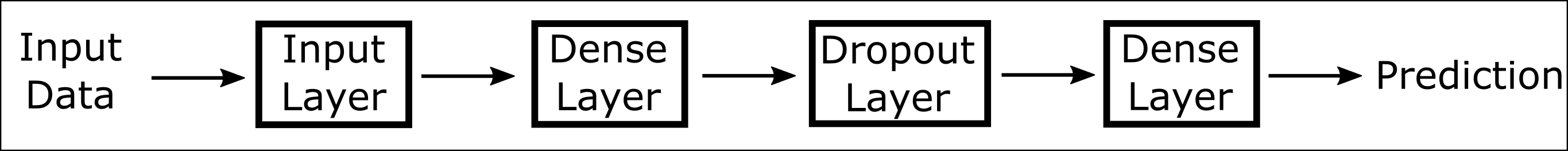}
    \caption{Structure of model using MLP only}
    \label{fig:model_mlp_only}
\end{figure}

The structural layout of our first model using only MLPs is given in Figure~\ref{fig:model_mlp_only}. Note that in this model, no LSTMs were used. The MLP layers are represented by dense layers. The first dense layer learns the features of each input while the second dense layer is the classifier. The experimental results for this model appear in Table~\ref{tab:results_mlp_only}. For five families, this model performs reasonably well with average accuracy of~83.56\%. However, the accuracy drops significantly when more families are added.

\begin{table}[!htb]
\centering
\caption{Results for the MLP model}
\label{tab:results_mlp_only}
\resizebox{0.45\textwidth}{!}{
\begin{tabular}{@{}ccc@{}}
\toprule
\multicolumn{1}{l}{\textbf{\begin{tabular}[c]{@{}l@{}}Number of Unique\\ Families to Classify\end{tabular}}} & \textbf{\begin{tabular}[c]{@{}c@{}}Accuracy Per\\ Experiment (\%)\end{tabular}} & \textbf{\begin{tabular}[c]{@{}c@{}}Average\\ Accuracy (\%)\end{tabular}} \\ 
\midrule \midrule
\multicolumn{1}{c|}{\multirow{5}{*}{5}} & \multicolumn{1}{c|}{81.95} & \multirow{5}{*}{83.56} \\
\multicolumn{1}{c|}{} & \multicolumn{1}{c|}{84.08} &  \\
\multicolumn{1}{c|}{} & \multicolumn{1}{c|}{82.41} &  \\
\multicolumn{1}{c|}{} & \multicolumn{1}{c|}{85.14} &  \\
\multicolumn{1}{c|}{} & \multicolumn{1}{c|}{84.21} &  \\ \midrule
\multicolumn{1}{c|}{\multirow{5}{*}{10}} & \multicolumn{1}{c|}{56.31} & \multirow{5}{*}{57.50} \\
\multicolumn{1}{c|}{} & \multicolumn{1}{c|}{56.81} &  \\
\multicolumn{1}{c|}{} & \multicolumn{1}{c|}{59.40} &  \\
\multicolumn{1}{c|}{} & \multicolumn{1}{c|}{61.50} &  \\
\multicolumn{1}{c|}{} & \multicolumn{1}{c|}{53.48} &  \\ \midrule
\multicolumn{1}{c|}{\multirow{5}{*}{15}} & \multicolumn{1}{c|}{49.27} & \multirow{5}{*}{51.22} \\
\multicolumn{1}{c|}{} & \multicolumn{1}{c|}{53.18} &  \\
\multicolumn{1}{c|}{} & \multicolumn{1}{c|}{54.82} &  \\
\multicolumn{1}{c|}{} & \multicolumn{1}{c|}{54.54} &  \\
\multicolumn{1}{c|}{} & \multicolumn{1}{c|}{44.31} &  \\ \midrule
\multicolumn{1}{c|}{\multirow{5}{*}{20}} & \multicolumn{1}{c|}{53.83} & \multirow{5}{*}{50.48} \\
\multicolumn{1}{c|}{} & \multicolumn{1}{c|}{45.68} &  \\
\multicolumn{1}{c|}{} & \multicolumn{1}{c|}{52.92} &  \\
\multicolumn{1}{c|}{} & \multicolumn{1}{c|}{46.87} &  \\
\multicolumn{1}{c|}{} & \multicolumn{1}{c|}{53.08} &  \\ \bottomrule
\end{tabular}
}
\end{table}


\subsection{LSTM without Embedding}
\begin{figure}[!htb]
    \centering
    \includegraphics[scale=0.3]{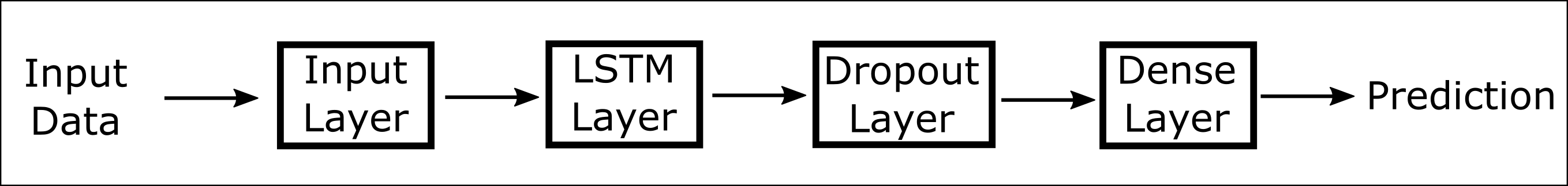}
    \caption{Structure of LSTM model without embedding}
    \label{fig:model_no_embedding}
\end{figure}

The structural layout of our basic LSTM model given in Figure~\ref{fig:model_no_embedding}.
Note that the model consists of four types layers, namely, an input layer, dropout layers, 
an LSTM layer, and a dense layer. The experimental results for this model
appear in Table~\ref{tab: results_no_embedding}. This model struggles with classifying just 
five families, with an average accuracy of~55.73\%. The accuracy drops as more families are 
classified. Clearly, a more sophisticated model is required.

\begin{table}[!htb]
\centering
\caption{Results for LSTM without embedding}
\resizebox{0.45\textwidth}{!}{
\begin{tabular}{@{}ccc@{}}
\toprule
\textbf{\begin{tabular}[c]{@{}c@{}}Number of Unique\\ Families to Classify\end{tabular}} & \textbf{\begin{tabular}[c]{@{}c@{}}Accuracy Per\\ Experiment (\%)\end{tabular}} & \textbf{\begin{tabular}[c]{@{}c@{}}Average\\ Accuracy (\%)\end{tabular}} \\
\midrule \midrule
\multicolumn{1}{c|}{\multirow{5}{*}{5}} & \multicolumn{1}{c|}{63.91} & \multirow{5}{*}{55.73} \\
\multicolumn{1}{c|}{} & \multicolumn{1}{c|}{48.44} &  \\
\multicolumn{1}{c|}{} & \multicolumn{1}{c|}{63.65} &  \\
\multicolumn{1}{c|}{} & \multicolumn{1}{c|}{42.94} &  \\
\multicolumn{1}{c|}{} & \multicolumn{1}{c|}{60.73} &  \\ \midrule
\multicolumn{1}{c|}{\multirow{5}{*}{10}} & \multicolumn{1}{c|}{40.50} & \multirow{5}{*}{39.28} \\
\multicolumn{1}{c|}{} & \multicolumn{1}{c|}{36.96} &  \\
\multicolumn{1}{c|}{} & \multicolumn{1}{c|}{41.46} &  \\
\multicolumn{1}{c|}{} & \multicolumn{1}{c|}{43.75} &  \\
\multicolumn{1}{c|}{} & \multicolumn{1}{c|}{33.71} &  \\ \midrule
\multicolumn{1}{c|}{\multirow{5}{*}{15}} & \multicolumn{1}{c|}{32.65} & \multirow{5}{*}{34.47} \\
\multicolumn{1}{c|}{} & \multicolumn{1}{c|}{35.56} &  \\
\multicolumn{1}{c|}{} & \multicolumn{1}{c|}{32.34} &  \\
\multicolumn{1}{c|}{} & \multicolumn{1}{c|}{35.06} &  \\
\multicolumn{1}{c|}{} & \multicolumn{1}{c|}{36.46} &  \\ \midrule
\multicolumn{1}{c|}{\multirow{5}{*}{20}} & 
 \multicolumn{1}{c|}{34.25} & \multirow{5}{*}{30.55} \\
\multicolumn{1}{c|}{} & \multicolumn{1}{c|}{27.74} &  \\
\multicolumn{1}{c|}{} & \multicolumn{1}{c|}{30.42} &  \\
\multicolumn{1}{c|}{} & \multicolumn{1}{c|}{30.45} &  \\
\multicolumn{1}{c|}{} & \multicolumn{1}{c|}{29.88} &  \\ \bottomrule
\end{tabular}
}
\label{tab: results_no_embedding}
\end{table}

\subsection{LSTM with Embedding}
\begin{figure}[!htb]
    \centering
    \includegraphics[scale=0.325]{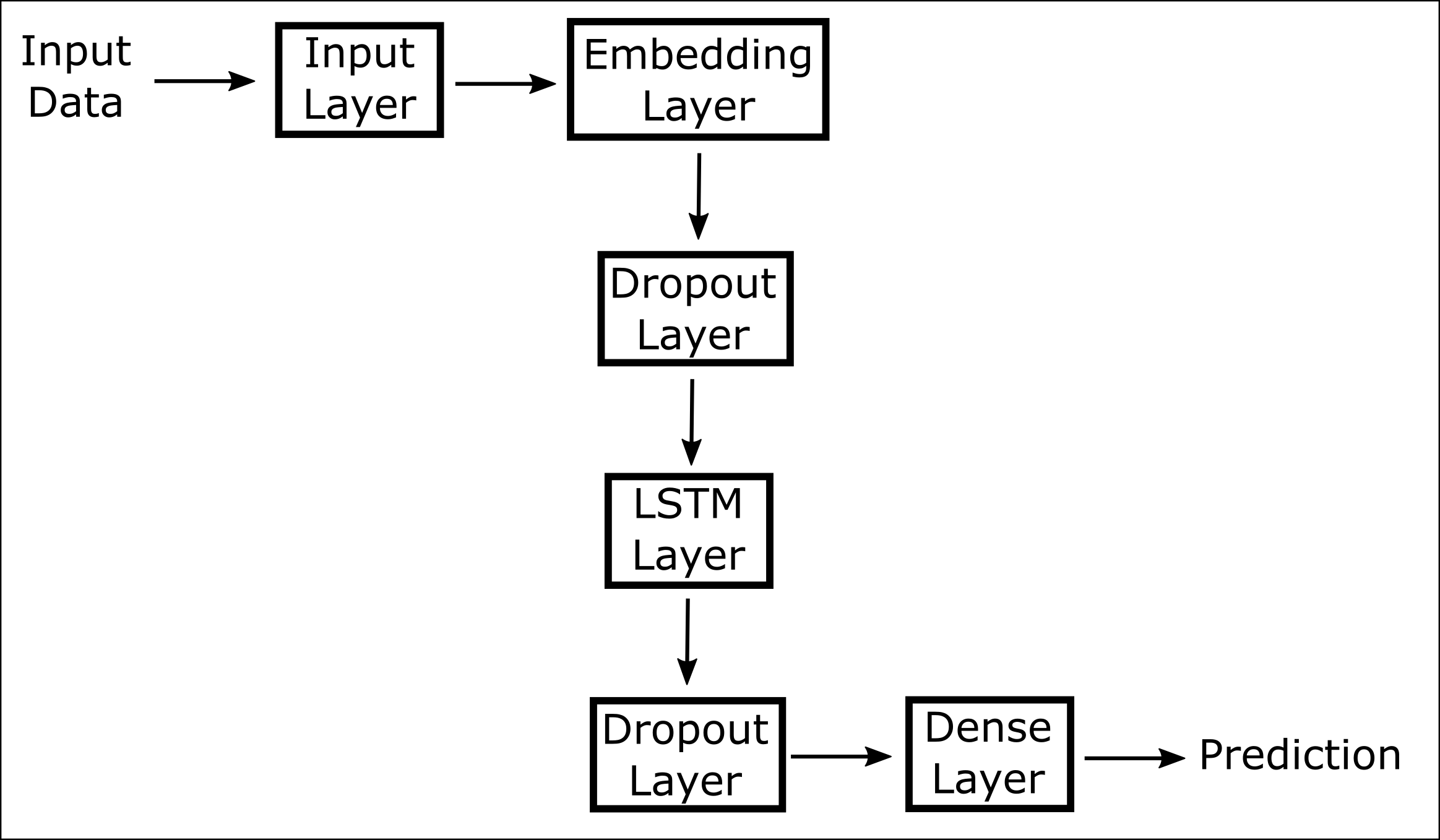}
    \caption{Structure of LSTM with embedding}
    \label{fig:model_with_embedding}
\end{figure}

In this model, we add an embedding layer to our basic LSTM,
as illustrated in Figure~\ref{fig:model_with_embedding}. Note that the embedding layer 
is between the input and LSTM layer. The experimental results for this model
are in Table~\ref{tab: results_with_embedding}. We see a significant improvement in 
the accuracy, with an average result of~74.66\%\ with 5 families, but
the accuracy drops dramatically when~10 or more families are considered.

\begin{table}[!htb]
\centering
\caption{Results for LSTM with embedding}
\resizebox{0.45\textwidth}{!}{
\begin{tabular}{@{}ccc@{}}
\toprule
\textbf{\begin{tabular}[c]{@{}c@{}}Number of Unique\\ Families to Classify\end{tabular}} & \textbf{\begin{tabular}[c]{@{}c@{}}Accuracy Per\\ Experiment (\%)\end{tabular}} & \textbf{\begin{tabular}[c]{@{}c@{}}Average\\ Accuracy (\%)\end{tabular}} \\
\midrule \midrule
\multicolumn{1}{c|}{\multirow{5}{*}{5}} & \multicolumn{1}{c|}{76.09} & \multirow{5}{*}{74.66} \\
\multicolumn{1}{c|}{} & \multicolumn{1}{c|}{73.64} &  \\
\multicolumn{1}{c|}{} & \multicolumn{1}{c|}{73.17} &  \\
\multicolumn{1}{c|}{} & \multicolumn{1}{c|}{76.90} &  \\
\multicolumn{1}{c|}{} & \multicolumn{1}{c|}{73.51} &  \\ \midrule
\multicolumn{1}{c|}{\multirow{5}{*}{10}} & \multicolumn{1}{c|}{54.46} & \multirow{5}{*}{54.89} \\
\multicolumn{1}{c|}{} & \multicolumn{1}{c|}{56.96} &  \\
\multicolumn{1}{c|}{} & \multicolumn{1}{c|}{55.67} &  \\
\multicolumn{1}{c|}{} & \multicolumn{1}{c|}{54.71} &  \\
\multicolumn{1}{c|}{} & \multicolumn{1}{c|}{52.90} &  \\ \midrule
\multicolumn{1}{c|}{\multirow{5}{*}{15}} & \multicolumn{1}{c|}{54.28} & \multirow{5}{*}{53.36} \\
\multicolumn{1}{c|}{} & \multicolumn{1}{c|}{51.97} &  \\
\multicolumn{1}{c|}{} & \multicolumn{1}{c|}{50.28} &  \\
\multicolumn{1}{c|}{} & \multicolumn{1}{c|}{53.22} &  \\
\multicolumn{1}{c|}{} & \multicolumn{1}{c|}{57.03} &  \\ \midrule
\multicolumn{1}{c|}{\multirow{5}{*}{20}} & \multicolumn{1}{c|}{51.11} & \multirow{5}{*}{49.66} \\
\multicolumn{1}{c|}{} & \multicolumn{1}{c|}{52.12} &  \\
\multicolumn{1}{c|}{} & \multicolumn{1}{c|}{51.60} &  \\
\multicolumn{1}{c|}{} & \multicolumn{1}{c|}{45.82} &  \\
\multicolumn{1}{c|}{} & \multicolumn{1}{c|}{47.65} &  \\ \bottomrule
\end{tabular}
}
\label{tab: results_with_embedding}
\end{table}

\subsection{BiLSTM with Embedding}
\begin{figure}[!htb]
    \centering
    \includegraphics[scale=0.325]{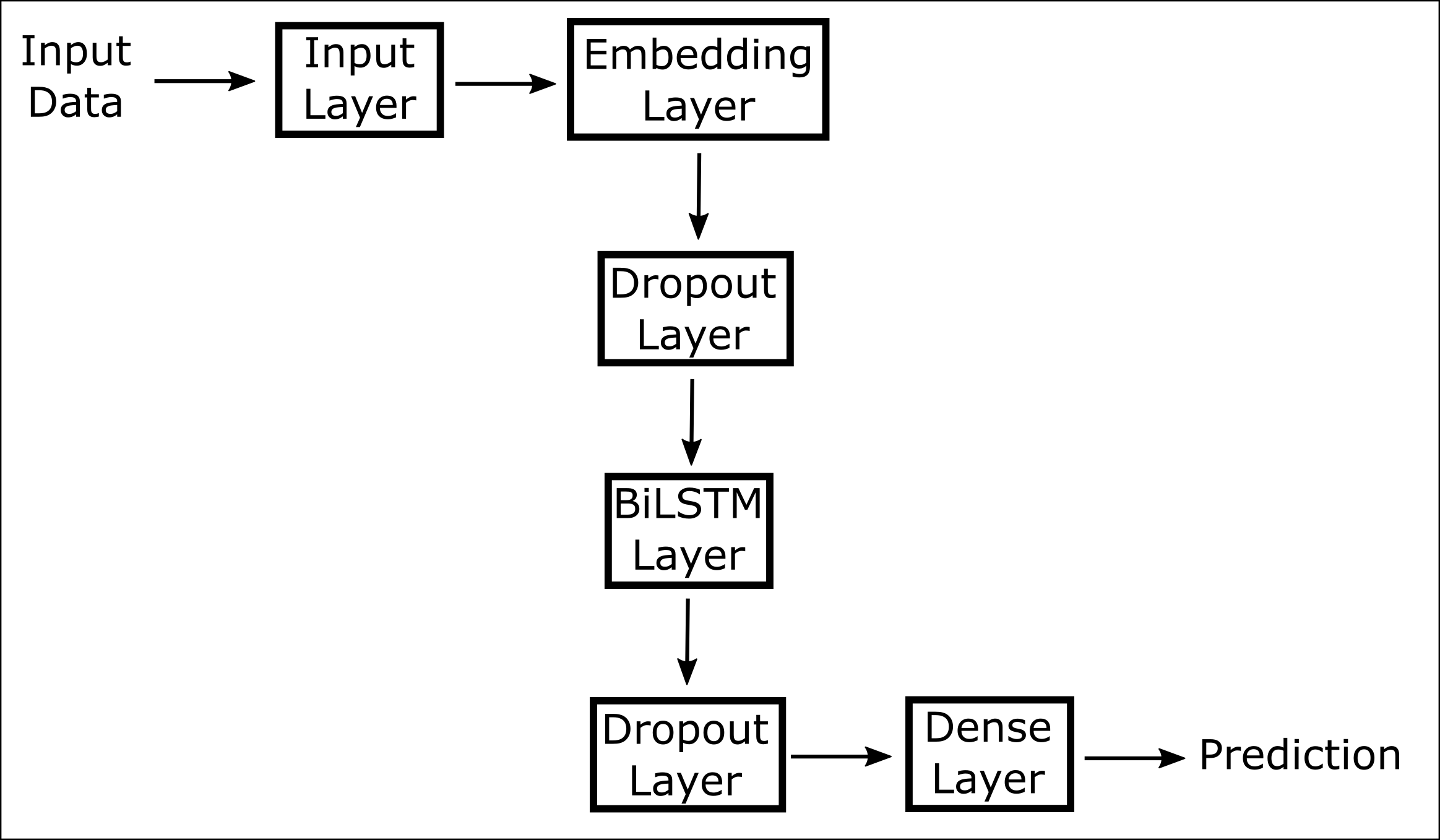}
    \caption{Structure of biLSTM with embedding}
    \label{fig:model_with_biLSTM}
\end{figure}

The structural layout of our first biLSTM model is shown in Figure~\ref{fig:model_with_biLSTM}. 
The only difference from our previous model is that the uni-directional LSTM layer 
has been replaced with a biLSTM layer. The experimental results for this model
are given in Table~\ref{tab: results_with_embedding}. From the results, we can see that 
a biLSTM is far more powerful than an LSTM in this context, as the accuracy 
has improved significantly. In fact, the accuracy when classifying~20 families with this biLSTM
model is nearly as good as the 5-family accuracy for the previous model. 

\begin{table}[!htb]
\centering
\caption{Results for biLSTM with embedding}
\resizebox{0.45\textwidth}{!}{
\begin{tabular}{@{}ccc@{}}
\toprule
\textbf{\begin{tabular}[c]{@{}c@{}}Number of Unique\\ Families to Classify\end{tabular}} & \textbf{\begin{tabular}[c]{@{}c@{}}Accuracy Per\\ Experiment (\%)\end{tabular}} & \textbf{\begin{tabular}[c]{@{}c@{}}Average\\ Accuracy (\%)\end{tabular}} \\ 
\midrule \midrule
\multicolumn{1}{c|}{\multirow{5}{*}{5}} & \multicolumn{1}{c|}{89.47} & \multirow{5}{*}{89.66} \\
\multicolumn{1}{c|}{} & \multicolumn{1}{c|}{90.83} &  \\
\multicolumn{1}{c|}{} & \multicolumn{1}{c|}{89.95} &  \\
\multicolumn{1}{c|}{} & \multicolumn{1}{c|}{85.94} &  \\
\multicolumn{1}{c|}{} & \multicolumn{1}{c|}{92.12} &  \\ \midrule
\multicolumn{1}{c|}{\multirow{5}{*}{10}} & \multicolumn{1}{c|}{79.58} & \multirow{5}{*}{79.30} \\
\multicolumn{1}{c|}{} & \multicolumn{1}{c|}{79.54} &  \\
\multicolumn{1}{c|}{} & \multicolumn{1}{c|}{78.13} &  \\
\multicolumn{1}{c|}{} & \multicolumn{1}{c|}{78.79} &  \\
\multicolumn{1}{c|}{} & \multicolumn{1}{c|}{80.46} &  \\ \midrule
\multicolumn{1}{c|}{\multirow{5}{*}{15}} & \multicolumn{1}{c|}{76.13} & \multirow{5}{*}{75.50} \\
\multicolumn{1}{c|}{} & \multicolumn{1}{c|}{76.13} &  \\
\multicolumn{1}{c|}{} & \multicolumn{1}{c|}{76.66} &  \\
\multicolumn{1}{c|}{} & \multicolumn{1}{c|}{76.28} &  \\
\multicolumn{1}{c|}{} & \multicolumn{1}{c|}{72.31} &  \\ \midrule
\multicolumn{1}{c|}{\multirow{5}{*}{20}} & \multicolumn{1}{c|}{73.71} & \multirow{5}{*}{73.36} \\
\multicolumn{1}{c|}{} & \multicolumn{1}{c|}{74.74} &  \\
\multicolumn{1}{c|}{} & \multicolumn{1}{c|}{69.53} &  \\
\multicolumn{1}{c|}{} & \multicolumn{1}{c|}{74.10} &  \\
\multicolumn{1}{c|}{} & \multicolumn{1}{c|}{74.72} &  \\ \bottomrule
\end{tabular}
}
\label{tab: results_with_embedding_bidirectional}
\end{table}

\subsection{BiLSTM with Embedding and CNN}
\begin{figure}[!htb]
    \centering
    \includegraphics[scale=0.25]{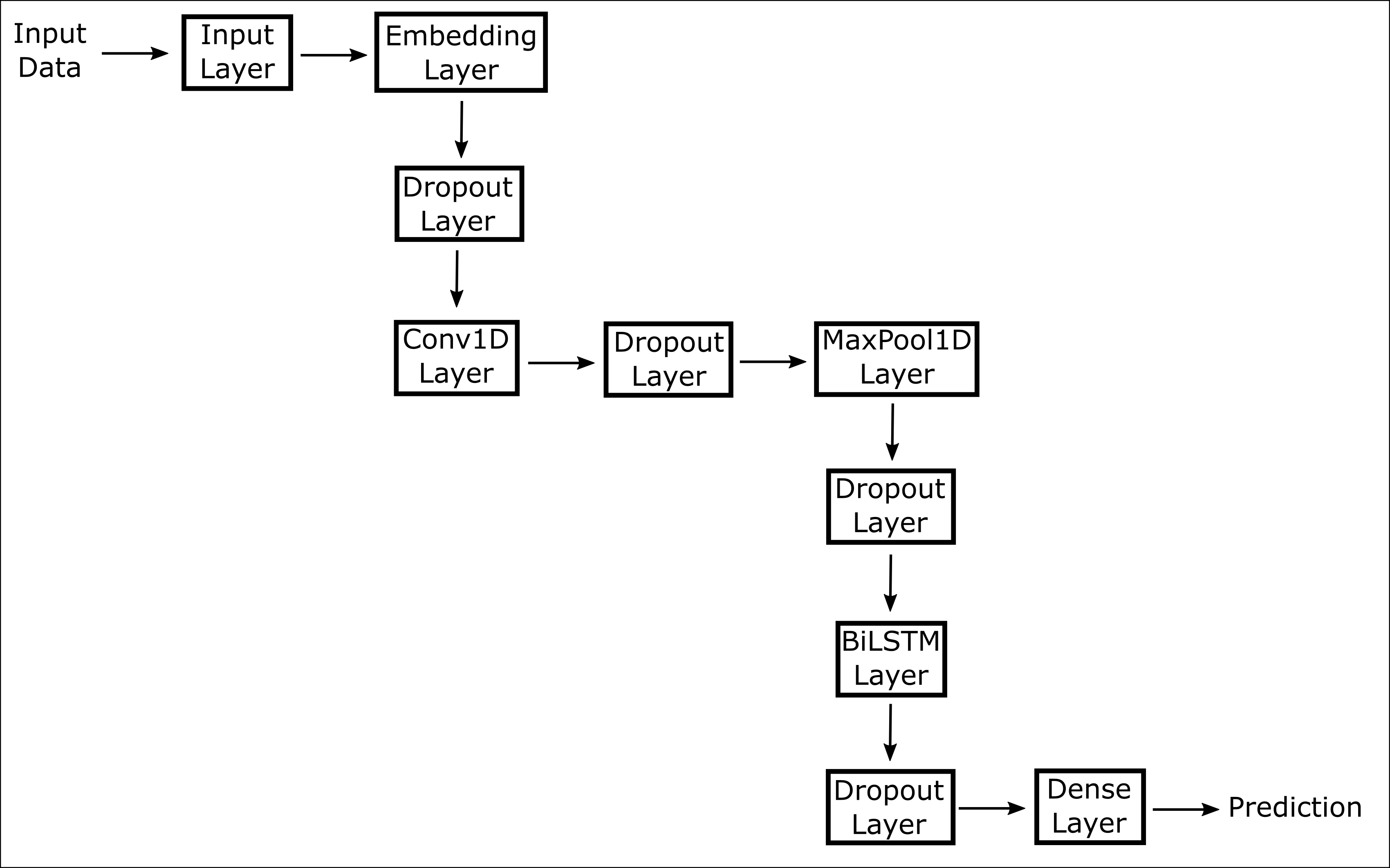}
    \caption{Structure of biLSTM, embedding, and CNN model}
    \label{fig:model_biLSTM_cnn}
\end{figure}

The structure of this model appears in Figure~\ref{fig:model_biLSTM_cnn}. 
Note that this model includes all of the layers as the previous model with the 
addition of a one-dimension convolutional layer and a max pooling layer. 
The experimental results for this case are given in Table~\ref{tab: results_with_embedding_bidirectional_cnn}. 
The addition of these CNN layers improves accuracy, and the improvement is
most significant as more families are considered---even for~20 families, we obtained 
a very respectable 81.06\%\ average accuracy. 

\begin{table}[!htb]
\centering
\caption{Results for biLSTM, embedding, and CNN model}
\resizebox{0.45\textwidth}{!}{
\begin{tabular}{@{}ccc@{}}
\toprule
\textbf{\begin{tabular}[c]{@{}c@{}}Number of Unique\\ Families to Classify\end{tabular}} & \textbf{\begin{tabular}[c]{@{}c@{}}Accuracy Per\\ Experiment (\%)\end{tabular}} & \textbf{\begin{tabular}[c]{@{}c@{}}Average\\ Accuracy (\%)\end{tabular}} \\ 
\midrule \midrule
\multicolumn{1}{c|}{\multirow{5}{*}{5}} & \multicolumn{1}{c|}{93.00} & \multirow{5}{*}{94.32} \\
\multicolumn{1}{c|}{} & \multicolumn{1}{c|}{96.33} &  \\
\multicolumn{1}{c|}{} & \multicolumn{1}{c|}{92.73} &  \\
\multicolumn{1}{c|}{} & \multicolumn{1}{c|}{94.70} &  \\
\multicolumn{1}{c|}{} & \multicolumn{1}{c|}{94.32} &  \\ \midrule
\multicolumn{1}{c|}{\multirow{5}{*}{10}} & \multicolumn{1}{c|}{90.42} & \multirow{5}{*}{87.38} \\
\multicolumn{1}{c|}{} & \multicolumn{1}{c|}{90.29} &  \\
\multicolumn{1}{c|}{} & \multicolumn{1}{c|}{81.29} &  \\
\multicolumn{1}{c|}{} & \multicolumn{1}{c|}{89.58} &  \\
\multicolumn{1}{c|}{} & \multicolumn{1}{c|}{85.29} &  \\ \midrule
\multicolumn{1}{c|}{\multirow{5}{*}{15}} & \multicolumn{1}{c|}{87.69} & \multirow{5}{*}{86.91} \\
\multicolumn{1}{c|}{} & \multicolumn{1}{c|}{87.56} &  \\
\multicolumn{1}{c|}{} & \multicolumn{1}{c|}{82.59} &  \\
\multicolumn{1}{c|}{} & \multicolumn{1}{c|}{87.31} &  \\
\multicolumn{1}{c|}{} & \multicolumn{1}{c|}{89.41} &  \\ \midrule
\multicolumn{1}{c|}{\multirow{5}{*}{20}} & \multicolumn{1}{c|}{83.29} & \multirow{5}{*}{81.06} \\
\multicolumn{1}{c|}{} & \multicolumn{1}{c|}{76.34} &  \\
\multicolumn{1}{c|}{} & \multicolumn{1}{c|}{80.60} &  \\
\multicolumn{1}{c|}{} & \multicolumn{1}{c|}{82.18} &  \\
\multicolumn{1}{c|}{} & \multicolumn{1}{c|}{82.88} &  \\ \bottomrule
\end{tabular}
}
\label{tab: results_with_embedding_bidirectional_cnn}
\end{table}

\subsection{Comparison of Results}
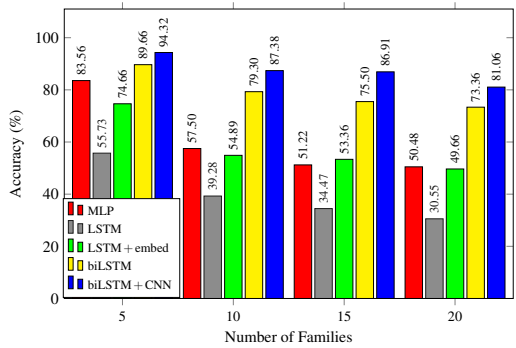
\begin{figure}[!htb]
\centering
  \begin{tikzpicture}[scale=0.6, every node/.style={scale=1.0}]
    \begin{axis}[
        width  = 0.8*\textwidth,
        height = 8cm,
        ymin=0,ymax=111,
        ytick={0,20,40,60,80,100},
        ybar=5*\pgflinewidth,
        bar width=11.0pt,
        xlabel = {Number of Families},
        ylabel = {Accuracy (\%)},
        symbolic x coords={5,10,15,20},
	y tick label style={
    		/pgf/number format/.cd,
   		fixed,
   		fixed zerofill,
    		precision=0},
        xtick={5,10,15,20},
        x tick label style={
		font=\small,
		},
        nodes near coords,
        every node near coord/.append style={rotate=90, 
        								   anchor=west,
								   font=\footnotesize,
								   /pgf/number format/.cd,
								   	fixed zerofill,
									precision=2
								   },
        enlarge x limits=0.175,
        legend cell align=left,
        legend pos=south west,
        legend style={
		font=\footnotesize,
                at={(0.1275,0.0025)},
                anchor=south,
                column sep=0.25ex
        }
    ]
\addplot[fill=red,opacity=1.00] 
coordinates {
(5,83.56)
(10,57.5)
(15,51.22)
(20,50.48)
};
\addplot[fill=gray,opacity=0.90] 
coordinates {
(5,55.73)
(10,39.28)
(15,34.47)
(20,30.55)
};
\addplot[fill=green,opacity=1.00] 
coordinates {
(5,74.66)
(10,54.89)
(15,53.36)
(20,49.66)
};
\addplot[fill=yellow,opacity=1.00] 
coordinates {
(5,89.66)
(10,79.30)
(15,75.50)
(20,73.36)
};
\addplot[fill=blue,opacity=1.00] 
coordinates {
(5,94.32)
(10,87.38)
(15,86.91)
(20,81.06)
};
\legend{MLP, LSTM, $\mbox{LSTM}+\mbox{embed}$, biLSTM, $\mbox{biLSTM}+\mbox{CNN}$}
\end{axis}
\end{tikzpicture}
\caption{Comparison of the average evaluation accuracy}
\label{fig:5_models_average_accuracy}
\end{figure}

A bar graph of the average accuracies for each model is shown in Figure~\ref{fig:5_models_average_accuracy}. 
As noted above, the basic LSTM model performs poorly, with each addition to the
model improving our results. 

The addition of an embedding layer dramatically increases the accuracy. This is not surprising, given
that previous work has shown that embedding layers can greatly improve the accuracy of
machine learning models applied to opcode sequences~\cite{aniket_paper}.

BiLSTMs and word embedding are often used together in NLP applications. However , their use in malware
research appears to be very uncommon to this point in time. Our models indicate that there is much
to be gained by considering both the forward and backward opcode sequence.

Finally, the addition of a one-dimensional CNN layer to the biLSTM and embedding layers 
gives the best performance among the four models studied in this research. Compared to the model 
without a CNN layer, the addition of this layer seems to have greater impact to performance when 
classifying more than~5 families. A possible explanation for why this model performs so well is 
that in addition to the benefits that come from having an embedding and biLSTM layers, a CNN 
layer helps the model by providing a different perspective on the opcode sequences. 
Specifically, CNNs focus the model on local structure whereas the biLSTM is focused on
overall characteristics. The interplay between these aspects---local and global---has the potential 
to provide the best of both, which we have married together into a single model.
The addition of a max pooling layer serves to further highlight the crucial aspects
of the local structure that the CNN highlights.

Confusion matrices for each model appear in the Appendix in
Figures~\ref{fig:confusion_matrix_no_embedding} 
through~\ref{fig:confusion_matrix_with_embedding_biLSTM_CNN}. 
These matrices show how often families are classified incorrectly and precisely where
these misclassifications occur. For example, considering our best model
results in Figure~\ref{fig:confusion_matrix_with_embedding_biLSTM_CNN},
we see that~4 families are badly misclassified, namely, Alureon, Obfuscator, Agent,
and Rbot, with, respectively, only~36\%, 31\%, 25\%, and~29\%\ classified
correctly. In contrast, 8 of the families are classified with~90\%\ or greater accuracy.

\section{Conclusion and Future Work}\label{sect:5}

In this research, we found that malware classification by by family using 
long-short term memory (LSTM) models is feasible. However, using just a single LSTM layer 
alone yields poor results. We found that by incorporating techniques from 
natural language processing (NLP), specifically, word embedding and bidirectional LSTMs (biLSTM), 
greatly improves the performance. We also discovered that that we could get obtain 
even better performance by including a convolutional neural network (CNN) layer in our model. 
Our best model was able to classify samples from~20 different malware families
with an average accuracy in excess of~81\%.
We conjecture that the interplay between the long-term memory of the biLSTM and
the local structure found by the CNN are the key to obtaining this strong performance.

For future work, more can be done into investigating why applying NLP techniques are 
so effective in classifying malware. The addition of an embedding layer, greatly improved 
our model's overall accuracy. Other techniques can be considered. For example, 
we might apply principle component analysis (PCA) 
to reduce the dimensionality of the weights obtained from the embedding layer. 
Additionally, experiments involving different word embedding algorithms (e.g., GloVe) 
would be worthwhile. Finally, further research into the possible benefits of combining LSTMs and CNNs
in this problem domain would be of great interest.

\bibliographystyle{apalike}

{\small
\bibliography{references.bib}
}

\section*{\uppercase{Appendix}}

\noindent Here, we provide confusion matrices for each of our experiments 
in Section~\ref{sect:4}.


\begin{figure}[!htb]
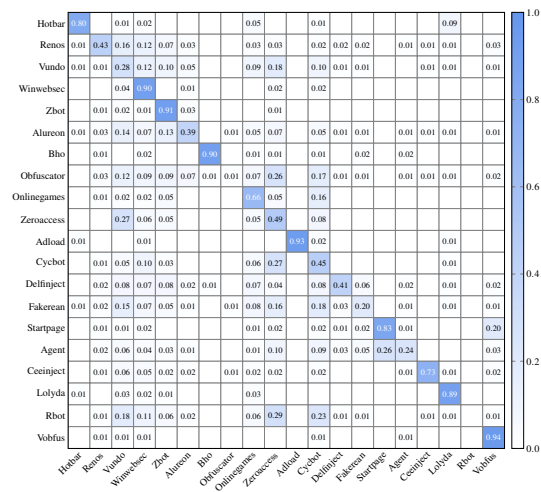

\centering
  \input figures/conf_mlp_only2.tex
    \caption{Confusion matrix for model using MLP only}
    \label{fig:confusion_matrix_mlp_only}
\end{figure}


\begin{figure}[!htb]
\centering
  \input figures/conf_no_embedding.tex
    \caption{Confusion matrix for LSTM without embedding}
    \label{fig:confusion_matrix_no_embedding}
\end{figure}


\begin{figure}[!htb]
\centering
  \input figures/conf_with_embedding.tex
    \caption{Confusion matrix for LSTM with embedding}
    \label{fig:confusion_matrix_with_embedding}
\end{figure}


\begin{figure}[!htb]
\centering
  \input figures/conf_embedding_biLSTM.tex
    \caption{Confusion matrix for biLSTM with embedding}
    \label{fig:confusion_matrix_with_embedding_biLSTM}
\end{figure}


\begin{figure}[!htb]
\centering
  \input figures/conf_biLSTM_CNN.tex
\caption{Confusion matrix for biLSTM with embedding and CNN}
\label{fig:confusion_matrix_with_embedding_biLSTM_CNN}
\end{figure}

\end{document}